\documentclass[preprint]{revtex4}
\usepackage{graphicx}
\usepackage{natbib}
\usepackage{url}
\usepackage{graphicx,amssymb,amsmath}

\begin{document}
\title{Diffusion Mechanisms in Li$_{0.5}$CoO$_2$ -- A Computational Study}

\author{Teut\"e Bunjaku}
\author{Andreas Pedersen}
\email{andped10@gmail.com}
\author{Mathieu Luisier}
\affiliation{Integrated Systems Laboratory, Department of Electrical
Engineering and Information Technology, ETH Zurich, Gloriastrasse
35, 8092 Zurich, Switzerland}

\begin{abstract}
An atomistic study of the order-effect occurring in Li$_{x}$CoO$_{2}$ at $x=0.5$ is presented and an explanation for the computationally and experimentally observed dip in the Li diffusivity is proposed.
Configurations where a single half-filled Li layer arranged in either a linear or a zig-zag pattern are simulated.
It is found that the lowest energy phase is the zig-zag pattern rather than the linear arrangement that currently is considered to be of lowest energy.
Atomic interactions are modeled at the DFT level of accuracy and energy barriers for Li-ion diffusion are determined from searches for first order saddle points on the resulting potential energy surface. 
The determined saddle points reveal that the barriers for diffusion parallel and perpendicular to the zig-zag phase differ significantly and explain the observed dip in diffusivity. 


\end{abstract}

\keywords{LiCoO$_2$, Diffusion, Saddle Point Searches}

\maketitle


\section{Introduction}
The diffusion coefficient of Li-ions is one of the key parameters in the design of high performance Li-ion batteries and defines a strict upper bound for how fast Li-ions can be inserted or extracted from the active material in the electrode. It thus limits the maximal current the battery can be operated at. Obtaining a battery with good charging and discharging characteristics requires therefore a high Li-ion diffusivity.


A widely used cathode material for Li-ion batteries is LiCoO$_2$. This material has been thoroughly examined using both experimental and modeling approaches. 
In an early computational study~\cite{VanderVen:1998ju} Van der Ven and Ceder determined that for a half filled LiCoO$_2$ layer the Li-ions preferably align themselves to form a linear phase. Other well ordered patterns, such as the zig-zag phase shown in Fig.~\ref{fig:phases}, are possible. However, the linear phase is currently considered to form as the Li-ion concentration reaches a coverage of 50\%. 

\begin{figure}
\includegraphics[width=0.7\textwidth]{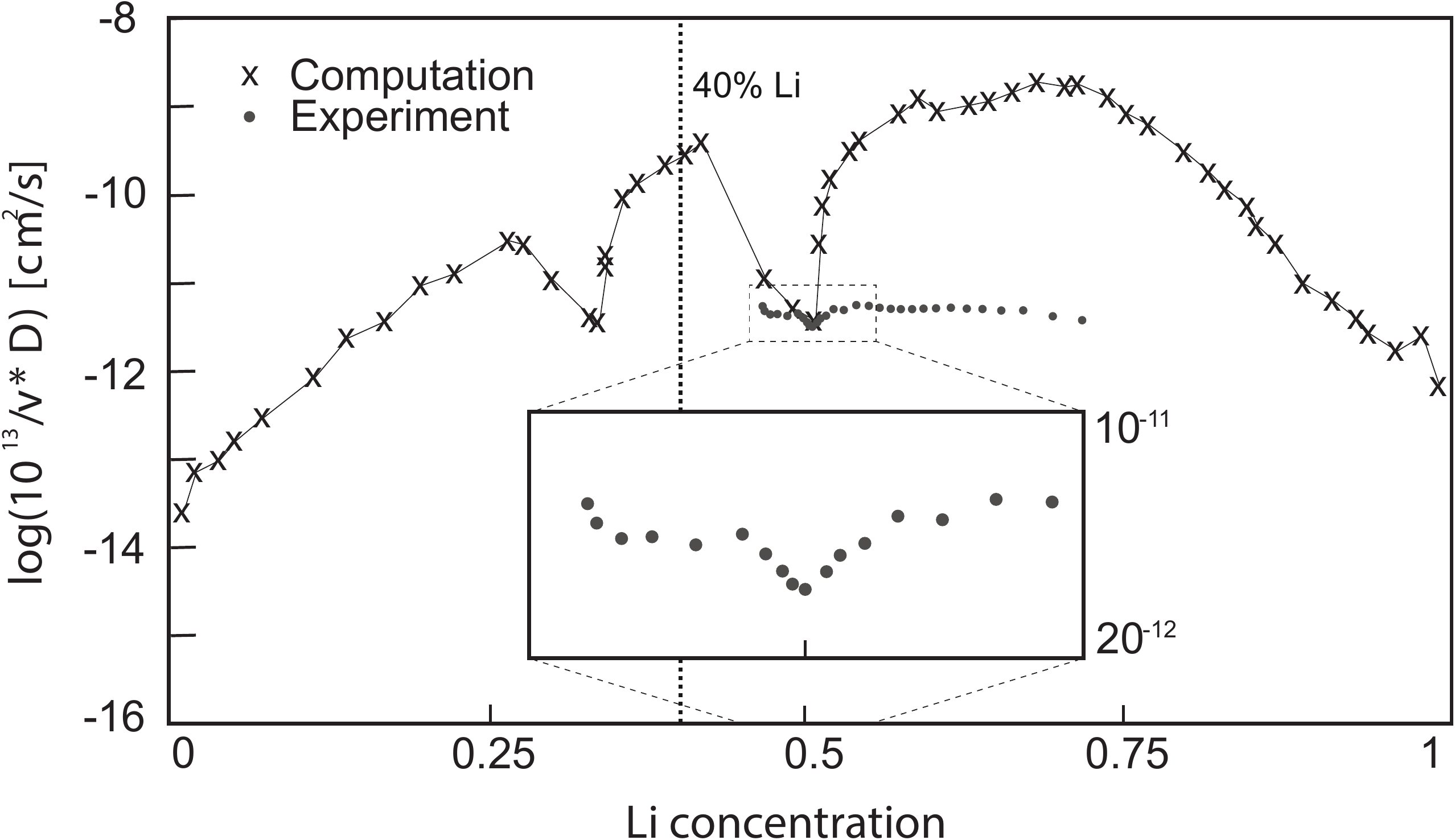}
\caption{Computational~\cite{VanderVen:2001ga} and experimental~\cite{Jang:2001ei} results of the Li-ion diffusion coefficient. A dip in the diffusivity at a half-filled level occurs in both computational and experimental investigations. The barrier for 40\% represents the disordered phase (dashed vertical line)}
\label{fig:dip}
\end{figure}

In a later work~\cite{VanderVen:2001ga} it was determined that the diffusive behavior of Li-ions between the stable CoO$_2$ layers is dominated by two mechanisms. The lowest barrier event was named 'tetrahedral site hop' ($\text{D}_{\text{TSH}}$) and  
involves a barrier varying from 0.3 to 0.6~eV depending on the configuration of the neighboring Li-ions. For the event to take place either a single or both lattice sites adjacent to the initial and final position of the moving Li-ion must be vacant. This mechanism dominates the low coverage regime.
The other event was named 'oxygen dumbbell hop' ($\text{D}_{\text{ODH}}$) where a Li-ion migrates to a vacant neighboring site using the shortest possible path. This mechanism dominates the high coverage regime and takes place when all adjacent sites are occupied. The barrier to be surpassed in this case is higher and is between 0.8 and 1.0~eV.
%
When considering the diffusion coefficient for Li-ions as a function of Li-load a strong dip occurs when the layers reach a half-filled level. Both computational~\cite{VanderVen:2001ga} and experimental~\cite{Jang:2001ei} results exhibit this feature that is shown in Fig.~\ref{fig:dip}. So far this dip has been attributed an abnormality of the ordered phase of the Li-ions. However, an in-depth explanation of the phenomenon is still lacking.

In the present work we show that the abrupt decrease in diffusivity is a consequence of an anisotropic behavior of Li-ions when they move parallel and perpendicular to the dominant pattern of the ordered phase. Furthermore, we demonstrate that the lowest energy configuration is the zig-zag phase, contrary to what has been considered so far. The energetic preference for this phase is explained by noticing that the zig-zag configuration enables a slight elongation of the nearest neighbor distances between the repelling Li-ions as compared to the linear phase.
These results are supplemented by simulations where first order saddle points (SPs) are located on the potential energy surface (PES) of the considered structures, as obtained from density-functional theory (DFT) calculations. From the determined SP it appears that the barriers for diffusion parallel and perpendicular to the zig-zag phase differ and this anisotropy explains the observed dip in diffusivity. In the next section we explained the applied computational methods. The key findings are then listed in the Results section and discussed while conclusions are drawn in the last section.

\section{Methodology}
\subsection{Structural Models}
The bulk phase of LiCoO$_2$ consists of stable cobalt oxide planes separating lithium layers that can be reversibly loaded and unloaded, as indicated in Fig.~\ref{fig:structure}. In a fully charged layer the Li-ions form a hexagonal pattern where each ion neighbors six others. In the present work the focus is on configurations where a Li-layer is only half full. For this coverage two ordered low energy phases have been determined, which are the linear phase, where the Li atoms align themselves in horizontal lines in the plane, and the zig-zag phase where the Li atoms form a zig-zag pattern, see Fig.~\ref{fig:phases}.
Both these phases are simulated to investigate how the diffusion barriers for the charge carrying Li-ions are affected by changes in the order of the phase. In both cases periodic boundary conditions along the $X$, $Y$, and $Z$ directions are applied to mimic bulk. The considered supercell contains three layers for the storage of Li-ions where one is half full and the others are fully loaded.

\subsubsection{Linear Phase}
For the linear pattern the supercell consists of a total of 553 atoms (120+1 Li, 144 Co, and 288 O atoms) decomposed into two fully filled Li layers with 48 Li atoms per layer, one half full layer of 24 atoms with an additional Li atom, where the additional Li atom is used to probe the diffusivity of the system, three Co layers, and six O layers. This system can be cast into a rectangular supercell with the dimensions $X=16.72$~\AA, $Y=19.31$~\AA, and $Z=13.58$~\AA. The cross section of the half full layer is shown in inset A of Fig.~\ref{fig:phases}.

\subsubsection{Zig-zag Phase}
For the zig-zag pattern the supercell is half the size of the one describing the linear phase. Its cross section is represented in inset C of Fig.~\ref{fig:phases}.
It contains 277 atoms (60+1 Li, 72 Co, and 144 O atoms). The number of layers and the height of the structure remain the same as for the linear phase but the number of atoms per full layer is reduced to 24. The size along the $X$ direction also remains unchanged whereas $Y$ decreases to $9.66$~\AA\ and becomes half the size of the supercell corresponding to the linear pattern. This reduced size suffices to model the zig-zag phase because the moving Li-ion in this case does not significantly interact with its periodically repeated image. A calculation using either $Y=9.66$~\AA\ or $Y=19.31$~\AA\ confirmed that the difference in the diffusion barrier does not exceed 0.01~eV, which here is considered to be insignificant.

\subsection{Atomic Interactions}
Energy and forces are computed using DFT as implemented in VASP~\cite{Kresse:1996vk,Kresse:1996kl}. The DFT calculations are undertaken at different levels of accuracy as two classes of simulations are made. An initial and comprehensive sampling of the possible mechanisms is done at low accuracy, which is followed by a series of refinement steps at higher accuracy. 
For the low accuracy simulations computational efficiency is of highest priority and the electronic degrees of freedom are restricted to the $\Gamma$-point. The exchange correlation functional applied is local density approximation and the cut-off energy for the planewave basis is 300~eV. Convergence is considered to be achieved when the energy difference between two consecutive iterations in the determination of the self-consistent field decreases below $1 \cdot 10^{-6}$~eV. For the higher accuracy refinement calculations the resolution of the k-point sampling sheme is enhanced to a 2x2x1 mesh on a Monkhorst-Pack grid.

\subsection{Saddle Point Determination}
Atomistic rearrangements of solid materials typically belong to the microsecond timescale. Such events are rare as compared to the period of lattice vibrations, which makes direct simulations relying on a standard scheme evolving Newton's law of motion an infeasible computational task.
To reach the long timescale behavior of atomistic systems more advanced methods are required. In the present case the dynamics is described by performing searches for saddle points using either the minimum mode following (MMF) or the nudged elastic band (NEB) method. By locating a SP on the potential energy surface (PES) one can determine the corresponding displacements that atoms must undergo to reach the connected product state. The corresponding process rate can be estimated from the energy of the SP configuration and the initial minimum. Within the harmonic approximation transition state theory (HTST) this rate, $k^{HTST}$, is given by: 
\begin{align}
\label{htst}
k^{HTST}\propto exp\left(-\frac{E_{SP}-E_{min}}{k_B T}\right),
\end{align}
where $E_{SP}$ is the energy of the SP configuration, $E_{min}$ the initial minimum, $k_B$ the Boltzmann constant, and $T$ the temperature.  This expression reveals that at moderate temperatures only low laying SPs are of importance since higher energy SPs only marginally contribute to the overall atomic displacement due to the exponential dependency of the rate equation. 

\subsubsection{Minimum Mode Following}
To gather an extensive collection of possible Li-ion diffusion mechanisms single ended searches for first order SPs are performed based on the MMF method~\cite{Henkelman:1999vr,Pedersen:2011cd} as implemented in EON~\cite{Pedersen:2010js,Chill:2014ct}. This approach has already been successfully applied to metallic~\cite{pedersen09_073034, pedersen09_4036, Benediktsson:2013hx}, covalent~\cite{Pizzagalli:2008gy, Pedersen:2009gv}, and molecular systems~\cite{Karssemeijer:2012ce, Pedersen:2014fo}. 
The searches for SPs are started from either of the two ordered phases in which a single additional Li-atom is inserted. To determine more complex mechanisms a series of searches is undertaken to resolve the full sequence of consecutive events. The search for SPs on a PES defined by DFT is achieved by forwarding the forces and energy determined by VASP to EON.
When applying the MMF method the vicinity of a SP on the PES is mapped onto a local minimum. Well-known minimization algorithms can then be applied for an efficient convergence onto a first order SP. 
More precisely, the energy of a dimer on the PES that is centered at the current configuration of the system is first minimized through rotation. This transformation orients the dimer approximately parallel to the lowest eigenvalue mode of the Hessian matrix and serves as an estimate for its minimum mode. The force acting on the system, $\mathbf{F}$, is then manipulated by inverting the component parallel to the determined direction, $\mathbf{v}_{\textit{min}}$. The resulting effective force, $\mathbf{F}^{\textit{eff}}$, acts on the system as:
\begin{align}
\label{equ:mmfm1}
\mathbf{F}^{\textit{eff}}=\mathbf{F}-2(\mathbf{F}\cdot\mathbf{v}_{\textit{min}})\mathbf{v}_{\textit{min}}.
\end{align}
By using this effective force, $\mathbf{F}^{\textit{eff}}$, an ordinary minimization method automatically converges onto first order SPs. In the current work L-BFGS is applied as the minimizer and convergence is achieved when the largest force acting on any atom decreases to a value smaller than 0.1~eV/\AA. 

Each search is initiated through the displacement of a single Li atom for which random components are added to its $X$, $Y$, and $Z$ coordinates. The displacement vector is drawn from a Gaussian distribution having 0.25~\AA\ as its standard deviation. In most cases the initial displacement occurs on the additional Li atom in the half filled Li layer. 
To accelerate the escape from regions close to a minimum where the lowest eigenvalue is positive, $\lambda_{\textit{min}}$, a more aggressive effective force is used:
\begin{align}
\label{equ:mmfm2} \mathbf{F}^{\textit{eff}}=
\begin{cases}
  -(\mathbf{F}\cdot\mathbf{v}_{\textit{min}})\mathbf{v}_{\textit{min}},& \text{ if \: $\lambda_{\textit{min}}>0$,} \\
  \mathbf{F}-2(\mathbf{F}\cdot\mathbf{v}_{\textit{min}})\mathbf{v}_{\textit{min}},& \text{ if \: $\lambda_{\textit{min}}<0.$}
\end{cases}
\end{align}
Additionally the bowl breakout~\cite{Pedersen:2014dg} extension to MMF is applied to further speed up the process of escaping the vicinity of a minima on the PES. This scheme confines the imposed displacements to the limited subset of $n_{bowl}=10$ atoms that are subject to the largest forces. Such confined displacements are then conducted within regions where all eigenvalues of the Hessian matrix are positive. Besides boosting the efficiency of reaching regions where one eigenvalue becomes negative it also ensures that the displacements keep a local character.

\subsubsection{Nudged Elastic Band}
Of the transition mechanisms determined in the MMF searches the lowest barrier events are recomputed with the nudged elastic band method~\cite{henkelman00_9901} to determine the minimum energy path (MEP). This additional step is done to determine the corresponding barriers with a higher accuracy model for the electronic degrees of freedom. These calculations are conducted using a climbing image scheme where the highest energy image is driven towards the SP by removing the spring force parallel to the band. The effective force on the top most image is constructed by inverting the real force components parallel to the tangent of the band. This tangent serves as an estimate for the direction of the lowest mode of the Hessian matrix similar to the dimer in the MMF scheme. 
To generate the initial guess for the band the bond conserving method~\cite{Smidstrup:2014bo} as implemented in the VNL software~\cite{www_vnl} is used. Depending on the complexity of the event a band of either 5 or 13 images is utilized in order to properly capture the diffusion mechanism. The ionic relaxation is consider converged when the largest force on any atom becomes smaller than 0.1~eV/\AA.

\section{Results}

The linear phase is characterized by an alignment of the Li-ions within the half filled layer. The resulting structure forms a highly ordered and symmetric pattern, detailed in inset A of Fig.~\ref{fig:phases}. It is considered to be the lowest energy configuration of Li$_{0.5}$CoO$_2$~\cite{VanderVen:1998ju}. Contrary to this previous finding the here presented results predict that the zig-zag phase exhibits a lower energy configuration. A lengthening  of the distances between the positively charge ions and thereby reduced repulsion forces offer an explanation for  this behaviour. This change is so subtle that its influence on the possible Li-ion diffusion mechanisms is insignificant. These quantities are discussed in the following sections.

\subsection{Linear Phase}
\begin{figure}
\includegraphics[width=0.55\textwidth]{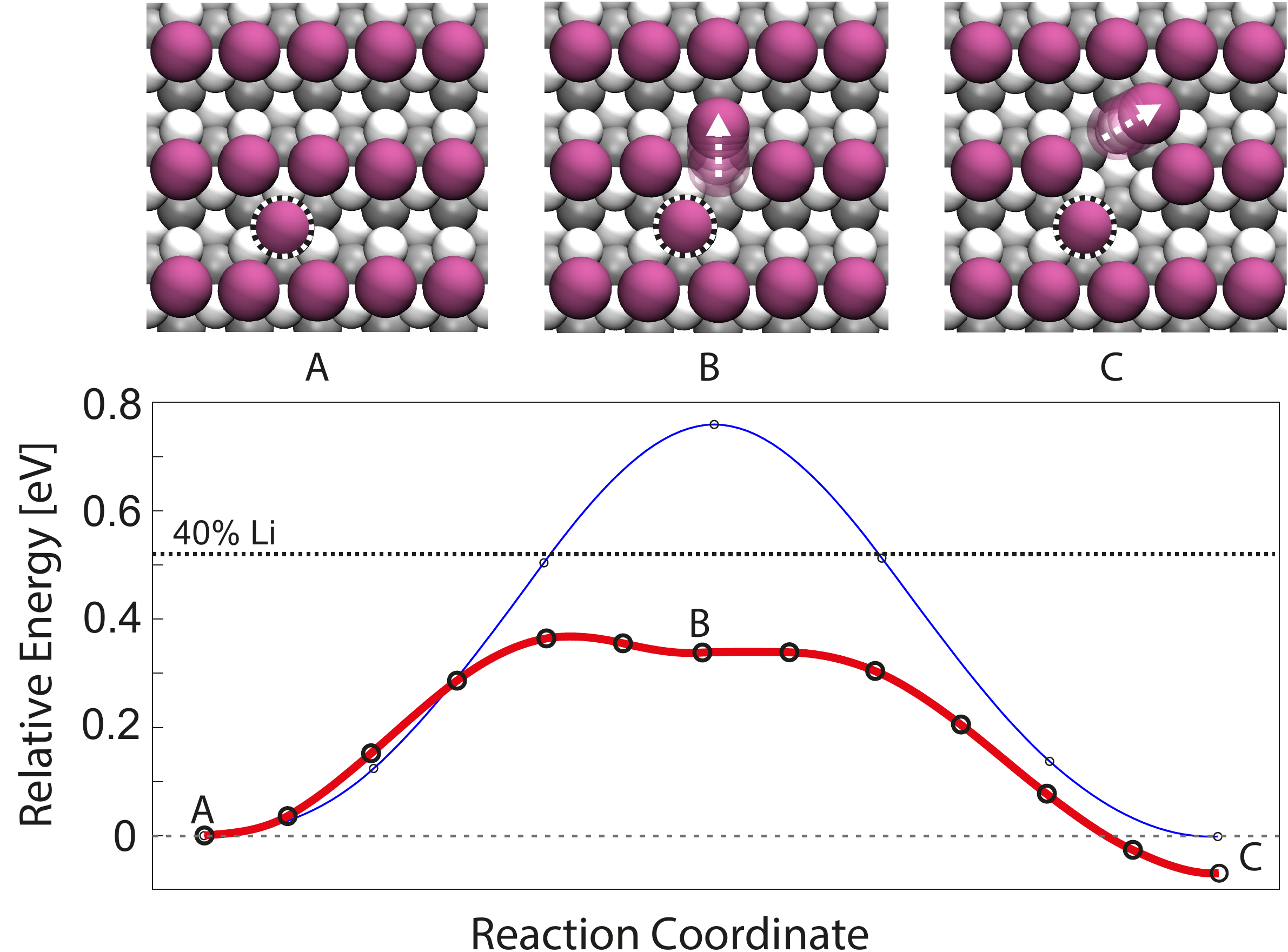}
\caption{Lowest barrier mechanism for the perpendicular diffusion in the linear phase. The barrier for this event is 0.36~eV (bold solid line) which is 0.16~eV lower than in the disordered phase (dashed horizontal line). The thin solid line represents the parallel diffusion. The perpendicular mechanism consists of two consecutive events. It is worthwhile noticing that the final state C is of slightly lower energy than the initial one (A). To reach a configuration where the line reforms the system must perform an event that is similar to A$\rightarrow$B.}
\label{fig:linear_perpendiular}
\end{figure}

To determine low barrier mechanisms for Li-ion diffusion a total of 40 MMF searches are conducted following the insertion of an additional Li atom. The lowest barrier mechanisms for a direct parallel and perpendicular displacement of a Li-ion and the corresponding MEPs are shown in Figs.~\ref{fig:linear_parallel} and \ref{fig:linear_perpendiular}.
%
%
From the MEP and the atomic dynamics of the parallel displacement it is clear that the lowest barrier mechanism only consist of a single diffusive event. The energy barrier for such a displacement to take place reaches 0.76~eV, which is 0.24~eV higher than the determined diffusion barrier for a configuration with a 40\% filling of the Li  
layer. This value is determined as an average from 20 MMF searches (avg = 0.52~eV, max = 0.54~eV, 
min = 0.51~eV).
When compared to earlier computational findings~\cite{VanderVen:2001ga} it appears that the here determined diffusion mechanism for a direct parallel displacement is similar to the already known $\text{D}_{\text{ODH}}$ event that is considered to involve a relatively high barrier. 

The perpendicular displacement occurs as two consecutive diffusion steps. Initially, the inserted Li-ion (atom) exerts a radial outward force on its four neighboring Li-ions until one of them leaves its site and breaks the symmetric linear arrangement as depicted in Fig.~\ref{fig:linear_perpendiular} A. For this to happen a 0.36~eV barrier must be surpassed. A meta stable state B is then reached before a second event takes place. The latter only involves a very low barrier of 0.4~meV and the Li-ion, which was pushed out from the line, then makes a horizontal move to reach the symmetric configuration C. 
To reform the linear ordered configuration two additional steps are required before the inserted Li-ion enters the vacant site in the neighboring line of Li-ions. 
These two consecutive steps are a mirrored version of the mechanism that broke the linear arrangement. A parallel movement of the Li-ion is performed first, followed by a perpendicular displacement that reforms the linear arrangement. The barrier for  regaining a linear and ordered structure is thus comparable to the barrier needed to break the initial linearity. It is however slightly higher due to the lower energy of the intermediate configuration C as compared to A.
This series of four $\text{D}_{\text{TSH}}$-like events represents the diffusion mechanism for a perpendicular displacement of a Li-ion. The highest barrier to circumvent as compared to the initial configuration, A, is 0.36~eV, which is 0.40~eV less than the barrier for the parallel path and 0.16~eV less than the 40\% filled Li layer. 

During the perpendicular displacement one of the Li-ions also undergoes a parallel displacement. This means that by conducting one upward and one downward move the system is able to make an indirect parallel displacement of a Li-ion by doing two $\text{D}_{\text{TSH}}$-like events rather than a direct displacement through the higher barrier $\text{D}_{\text{ODH}}$ event.

\subsection{Zig-zag Phase}
\begin{figure}
\includegraphics[width=0.5\textwidth]{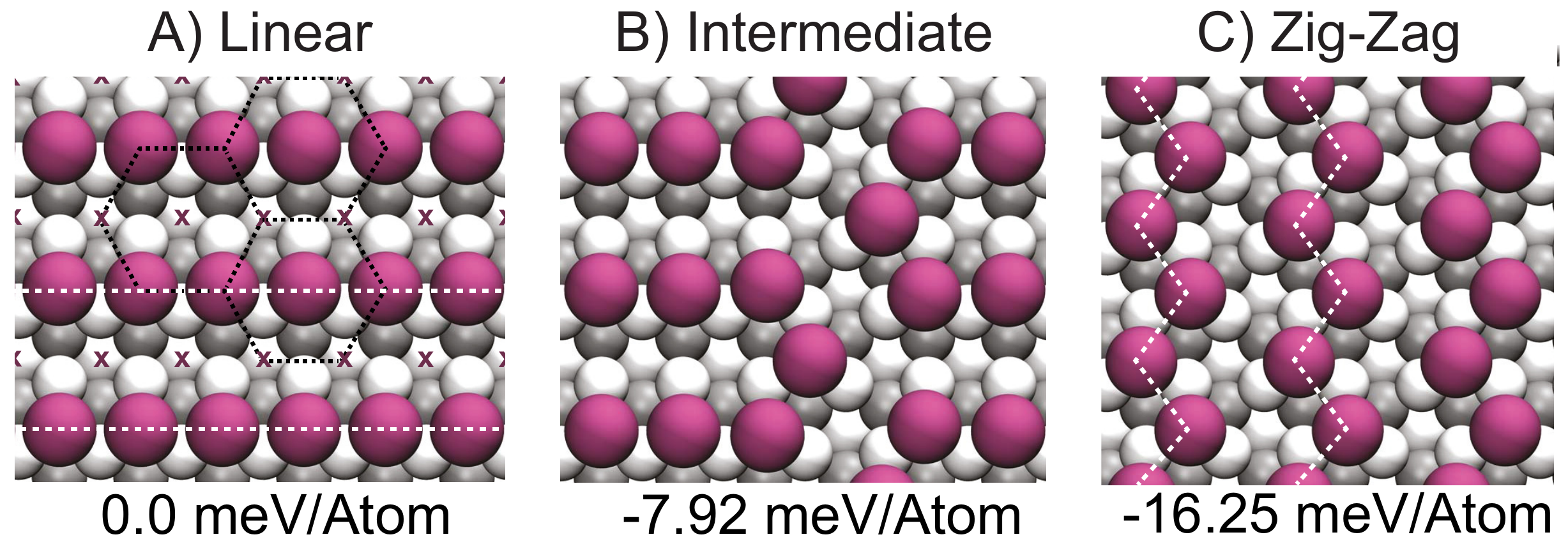}
\caption{Cross sections showing different phases of a half full Li-layer in LiCoO$_2$. 
Inset A: linear phase marked by the dashed white lines. The lattice sites of missing Li atoms are indicated by an x. Each Li atom would have six neighbors in a fully loaded layer, thus forming a hexagonal structure as marked by the black dashed lines. 
Inset B: cross section showing a partical zig-zag phase realized after several perpendicular diffusion steps are performed in the linear pattern. This structural change causes an energy lowering by 7.92~meV/Atom as compared to the linear phase. 
Inset C: full transformation to the zig-zag phase that results in a further lowering of the energy by 16.25meV/Atom.}
\label{fig:phases}
\end{figure}

\begin{figure}
\includegraphics[width=0.3\textwidth]{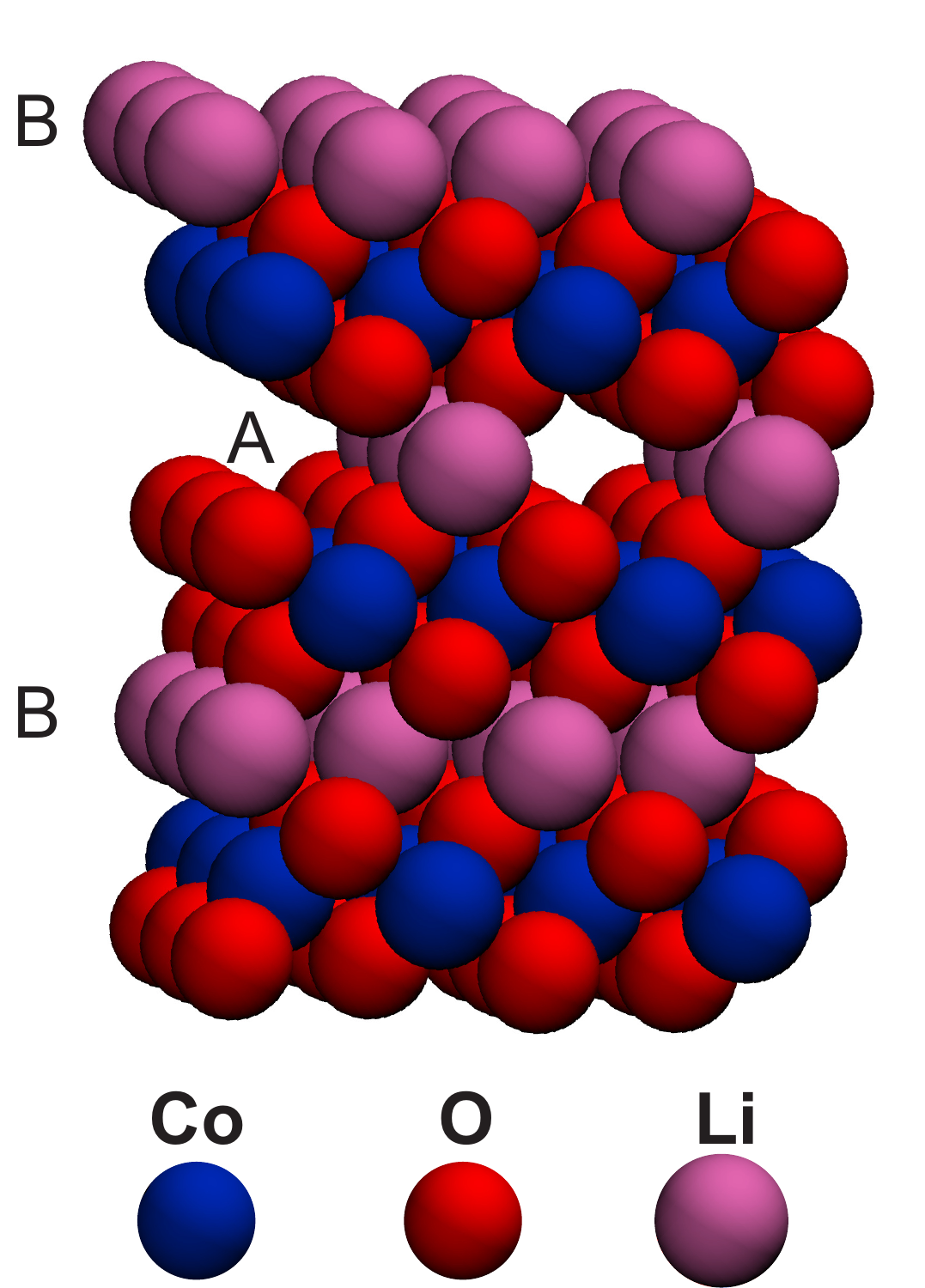}
\caption{Li$_x$CoO$_2$ atomic arrangement. In the used structure two of three Li intercalation layers are fully loaded (B) whereas the Li coverage of the remaining layer (A) is either $x=0.4$ or $x=0.5$ (A)
For the shown structure $x$ is 0.5 and the ions are arranged in the linear phase introduced in Fig.~\ref{fig:phases}.}
\label{fig:structure}
\end{figure}

The zig-zag phase results as the lowest energy configuration after several perpendicular displacements are imposed on the linear phase. This transformation is shown in Fig.~\ref{fig:phases}, which starts from the linear phase, goes through an intermediate step, before the final and lowest energy zig-zag phase is obtained. In the initial linear configuration, inset A, the distance between neighboring Li-ions in the half full layer is 2.79~\AA.
For the intermediate phase, inset B, the linear phase is terminated with a zig-zig pattern and an energy lowering of 7.92~meV/Atom is gained. The distances between the Li-ions in the termination sections have increased and are within an interval from 2.81 to 2.95\AA\ that explain the energy lowering.
The extended zig-zag pattern results from further rearrangements of the intermediate configuration, inset C in Fig.~\ref{fig:phases}, which is characterized by an extension of all Li-ion bonds in the half full layer. They now reach a length of 2.97~\AA\ and result in  a further energy decrease. The total energy lowering reaches 16.25~meV/Atom as compared to the linear phase. The discrepancy between this lowest energy phase and earlier studies~\cite{VanderVen:1998ju} might be explained by the fact that in the previous work a smaller unit cell was used due to the present limitations on computational resources. A too small cell size will not allow for the here observed reconstruction of the Li-ions as the system will lack the required ionic degrees of freedom.

%

\begin{figure}
\includegraphics[width=0.55\textwidth]{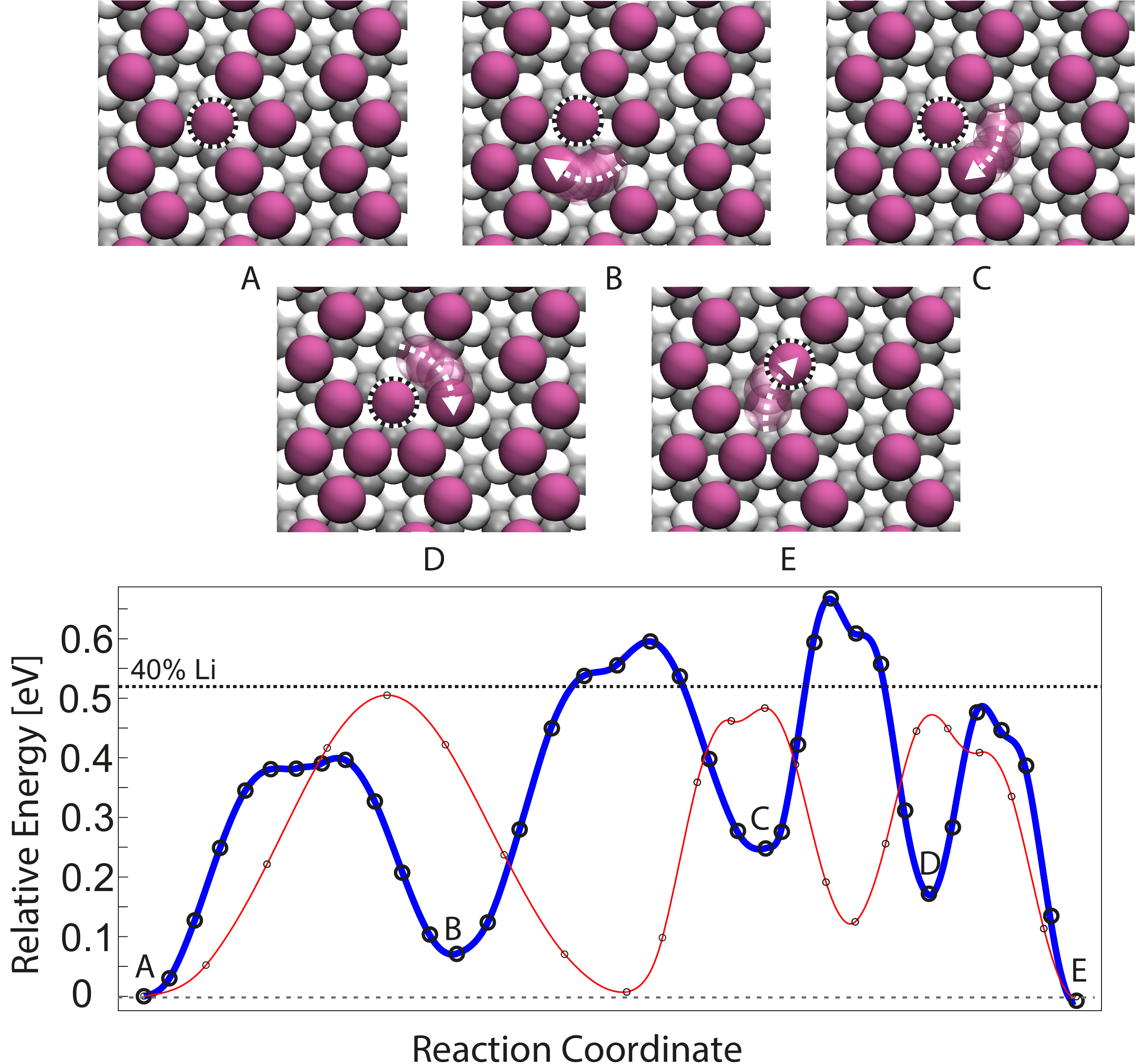}
\caption{Lowest barrier mechanism for parallel diffusion in the zig-zag phase. The highest barrier of the four required events shown with the bold solid line is 0.67~eV, which is 0.15~eV higher than within the disordered phase, dashed horizontal line. The minimum energy path for perpendicular diffusion is shown with a thin solid line. Subplots A to E show the visited metastable states and the end configurations.}
\label{fig:zigzag_parallel}
\end{figure}
\begin{figure}
\includegraphics[width=0.55\textwidth]{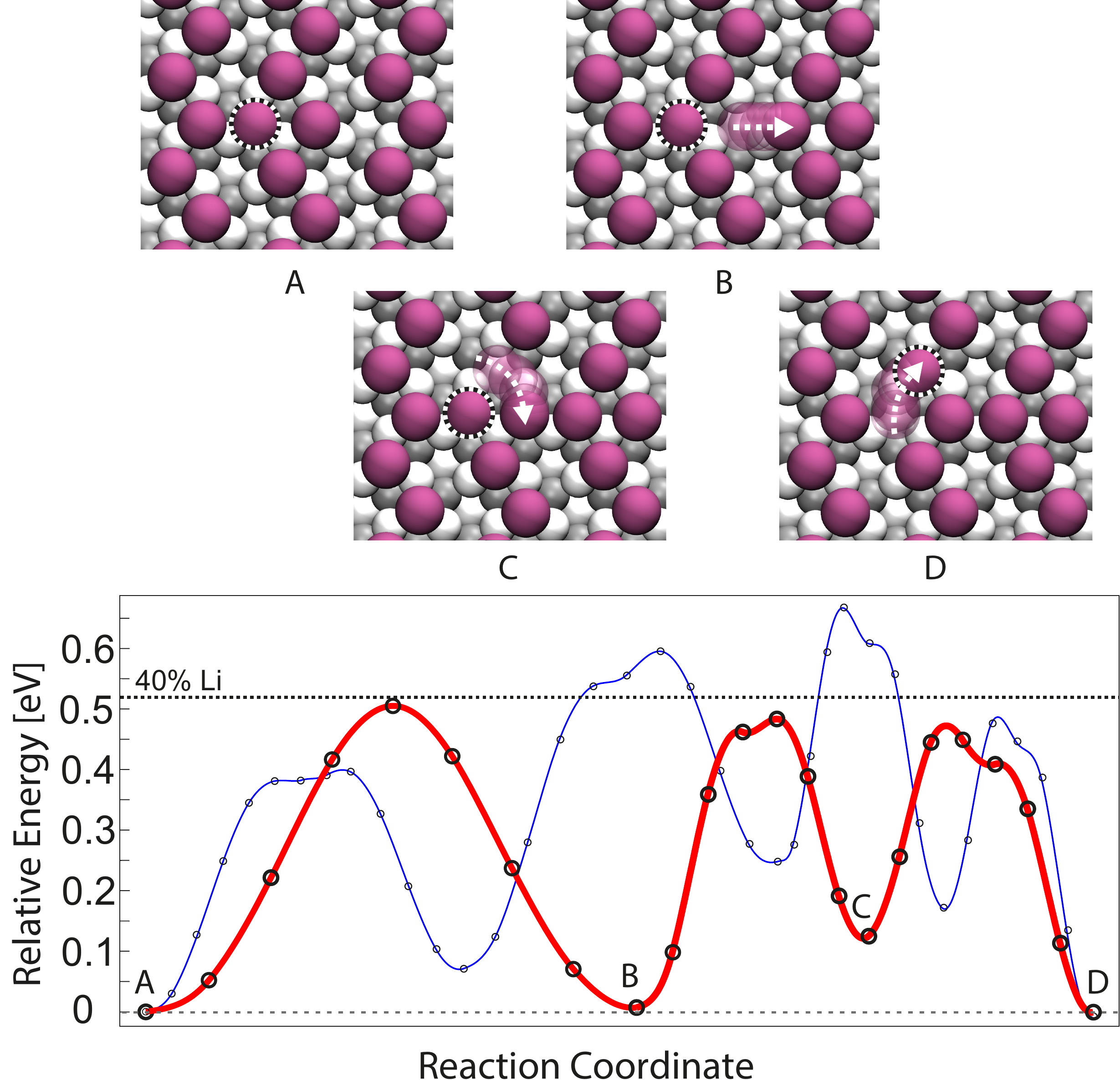}
\caption{Lowest barrier mechanism for perpendicular diffusion in the zig-zag phase. The barrier height for this mechanism is 0.51~eV and it involves three consecutive events shown with the bold solid line. As for the linear phase (Fig.~\ref{fig:linear_perpendiular}) it appears that the barrier for perpendicular diffusion is significantly smaller than for a parallel displacement (thin solid line) but similar or smaller than for the disordered phase that is marked by the dashed horizontal line.}
\label{fig:zigzag_perpendiular}
\end{figure}

Starting from the determined lowest energy configuration a second series of SP searches using MMF is undertaken. The two resulting lowest barrier mechanisms for parallel and perpendicular diffusion in the zig-zag phase are shown in Figs.~\ref{fig:zigzag_parallel} and \ref{fig:zigzag_perpendiular}. Similarly to the linear phase the parallel displacement consists of higher barrier events, whereas the barriers involved in the perpendicular mechanism are smaller.
%
The MEP for the parallel path, Fig.~\ref{fig:zigzag_parallel}, exhibits an activation barrier of 0.67~eV, which is significantly higher than the activation barrier of 0.52~eV determined for a 40\% filling of the Li layer. It also appears that the mechanism for a parallel displacement consists of four consecutive $\text{D}_{\text{TSH}}$ like steps. This is the most complex of the determined mechanisms. 
%
For perpendicular diffusion the barrier height is 0.51~eV, which is 0.16~eV lower than the parallel barrier and thus comparable to a 40\% filled layer. The displacement consists of three events and is a slightly more complex mechanism than for the linear phase where three consecutive steps are also required. The MEP and the corresponding atom configuration are shown in Fig. \ref{fig:zigzag_perpendiular}. In this case the series is three $\text{D}_{\text{TSH}}$-like diffusive events.



\section{Discussion}

%
%
The higher energy of the linear phase is strong evidence that the zig-zag phase is the true configuration for a half filled Li layer. Furthermore, the small activation barriers for both diffusion paths in the linear phase cannot explain the observed dip in diffusivity, which rather should show enhanced dynamics that is a direct contradiction of the observation for a half filled layer. 
On contrary the anisotropic diffusion in  the zig-zag phase could explain this behavior.
From the results for the zig-zag phase it appears that only for an ideal sample, with a single domain extending throughout the entire crystal, an efficient (un)loading of Li-ions could take place only  by undergoing perpendicular diffusion. However, any real sample will consist of many differently oriented domains that will force the Li-ions either to undergo the higher barrier horizontal moves and stay within the domain, or enter a less favorable site of higher energy at the domain boundary. The explanation for the dip in diffusivity is therefore that the Li-ions will be forced to undergo the less favorable parallel mechanisms and become the bottleneck for diffusion.



\begin{figure}
\includegraphics[width=0.55\textwidth]{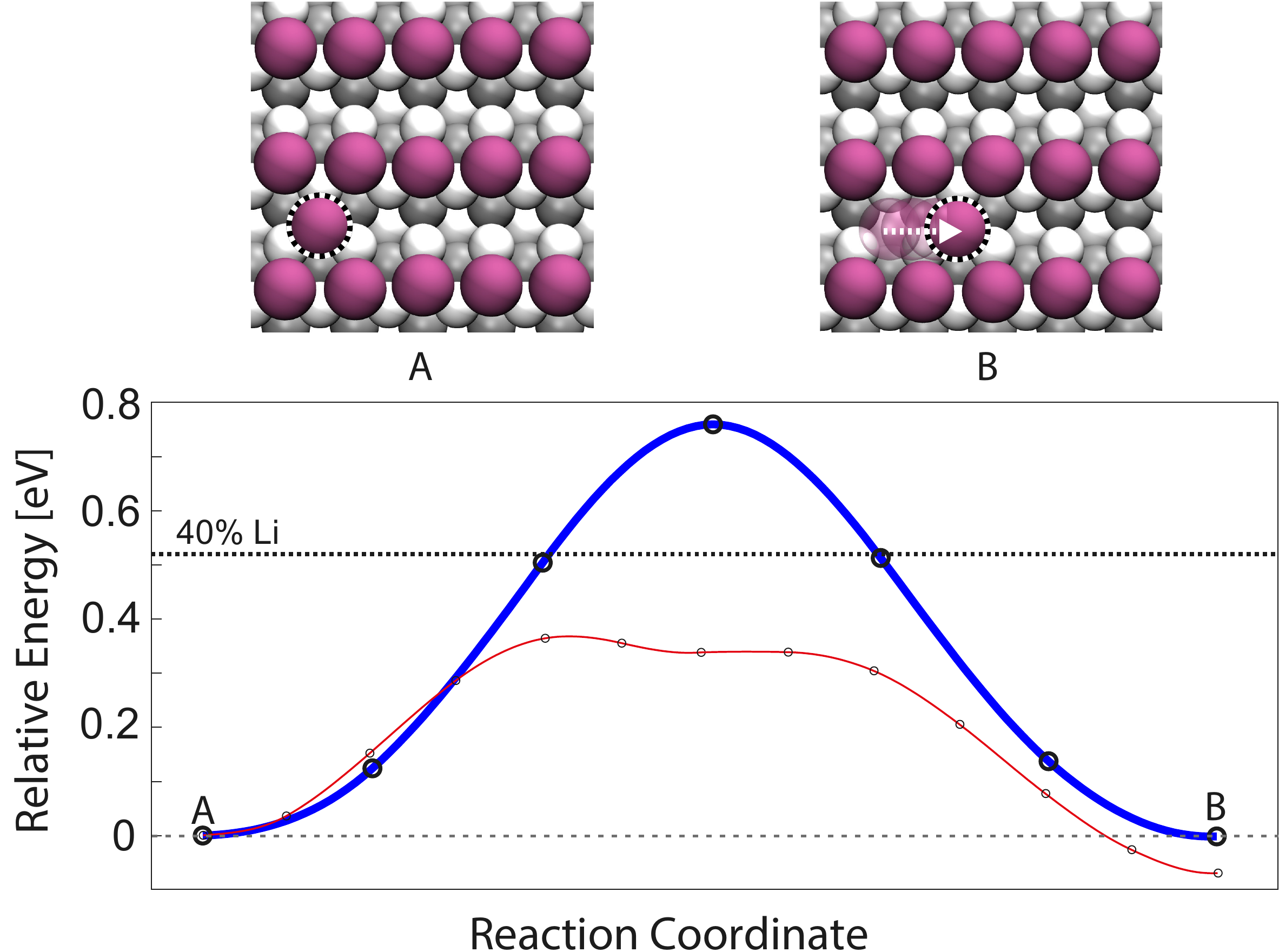}
\caption{Minimum energy path of the lowest barrier mechanism for parallel diffusion in the linear phase. The energy barrier is 0.76~eV (bold solid line) which is 0.24~eV higher than the barrier at 40\% representing the disordered phase (dashed horizontal line). The thin solid line refers to the perpendicular diffusion. The insets illustrate the displacement of the additional Li atom by one lattice site parallel to the lines formed by the other Li atoms. The stable configurations are shown. The arrow marks the atom that undergoes the most significant displacement.}
\label{fig:linear_parallel}
\end{figure}

The two basic mechanisms for Li-ion diffusion observed by Van der Ven and Ceder~\cite{VanderVen:2001ga} are recovered in the present work. 
The $\text{D}_{\text{ODH}}$ mechanism occurs during the parallel displacement in the linear phase, Fig.~\ref{fig:linear_parallel}, where the inserted Li-ion is forced to move between two Li-ions belonging to the linear pattern. Due to the repelling character of the charged ions this mechanism shows a high barrier because the displaced ion gets into a close vicinity of its neighbors before the final state is reached. The $\text{D}_{\text{TSH}}$ mechanism can be observed in the perpendicular move in the linear phase as depicted in Fig.~\ref{fig:linear_perpendiular}. Instead of moving between two Li-ions, the additional ion 'pushes' a neighbor out of its position. When breaking up a line the moving Li-ion does not have to come into the close vicinity of another ion and a lower energy barrier therefore results. 

When comparing the complexity of the diffusion mechanisms for the two phases, the ones occurring in the linear phase, Fig.~\ref{fig:linear_parallel} and \ref{fig:linear_perpendiular}, are the easiest to comprehended. For the linear phase the mechanisms only involve the inserted Li-ion and a single Li-ion from the existing pattern whereas displacements in the zig-zag phase, Fig.~\ref{fig:zigzag_parallel} and \ref{fig:zigzag_perpendiular}, require a rearrangement of several of the neighboring Li-ions. In all cases the inserted Li-ion is the last one to move, which appears to be a characteristic for the dynamics of repelling bodies. It is also noteworthy that the lowest barrier events only cause a significant displacement of a single Li-ion rather than concerted moves where several atoms are displaced, as found in metallic systems~\cite{pedersen09_073034}.  

For both phases the diffusion barriers for a direct parallel or perpendicular displacement to take place are significantly different. A movement parallel to the dominant pattern involves the highest energy barrier. However, in the linear phase an alternative, indirect route exists: parallel displacements can be achieved if the system undergoes two perpendicular moves. This eliminates the barrier anisotropy in the linear phase whereas it remains in the zig-zag phase.
The anisotropy in barrier heights combined with the fact that the zig-zag phase is lower in energy offer an explanation for the reduced diffusivity at $x=0.5$. Due to the difference in barrier heights the Li-ions are only able to diffuse perpendicularly to the ordered pattern. Parallel moves are inhibited and thereby hinder a homogenous loading/unloading. 
Diffusion in the zig-zag phase consists of $\text{D}_{\text{TSH}}$-like events, with a lower energy barrier when compared to the $\text{D}_{\text{ODH}}$ mechanism.
 Even though the diffusion path can be obtained by only traversing low barrier $\text{D}_{\text{TSH}}$ SPs the overall activation barrier is high. This is due to the disordered intermediate states B to D (see Fig.~\ref{fig:zigzag_parallel})
that have a higher energy compared to the states A and E, since for a half filled Li layer only the highly ordered zig-zag phase represents a stable configuration, whereas for the 40\% filled Li layer the 
intermediate
 states are often stable configurations as well. This increase in energy for the metastable states causes the SPs to reach a higher value and results in a higher barrier even though no $\text{D}_{\text{ODH}}$ like event needs to take place.

\section{Conclusion}

By using density functional theory and determining first order saddle points on the resulting potential energy surface we have explained the reason behind the experimentally and computationally observed dip in the diffusivity for half filled Li-layers in Li$_x$CoO$_2$ ($x$=0.5). Up to now this phenomenon has been attributed to an ordering effect, but from the present atomistic simulations we are able to demonstrate in great detail that it is due to an anisotropy in parallel and perpendicular diffusion in the zig-zag phase. Since diffusion parallel to the dominant pattern involves a higher barrier an efficient loading and unloading of Li-ions is hindered at $x$=0.5.  
Furthermore, we have shown that the lowest energy phase for a half full Li-layer in LiCoO$_2$ is a zig-zag pattern rather than a linear arrangement. This finding is backed by indirect experimental support since the anisotropy for diffusion, which causes the dip in diffusivity, does not exist in the linear phase.

\begin{acknowledgments}
This research is funded by the EU Commission (ERC starting
grant: E-MOBILE) and the platform for Advanced Scientific Computing in Switzerland (PASC). The computer simulations are done at the Swiss
National Supercomputer Center (projects: s579 and s591).
\end{acknowledgments}

\bibliographystyle{apsrev4-1}
\bibliography{LiCoO2,LiCoO2_web}

\end{document}